\documentclass[preprint]{acmart}
\AtBeginDocument{%
  }

\usepackage{booktabs}
\usepackage{array}

\begin{document}

\renewcommand\footnotetextcopyrightpermission[1]{}  

\settopmatter{printacmref=false, printccs=false, printfolios=true}

\title{Measuring What Matters: A Quantitative UX Evaluation Framework for AI-Assisted Home Search}

\author{Matilda Nkoom}
\email{matilda\_nkoom@tamu.edu}
\correspondingauthor
\authornotemark[1]
\affiliation{%
  \institution{Texas A\&M University}
  \city{College Station}
  \state{Texas}
  \country{USA}
}









\begin{abstract}
 AI-assisted conversational search is rapidly displacing filter-based interfaces across the major home search portals. Redfin's deployment of conversational search produced a 47\% lift in tour requests, and Zillow launched "AI Mode" in March 2026. Recent consumer surveys indicate that a large majority of Americans now use AI tools for housing market information. Yet the evaluation frameworks practitioners apply to these products remain borrowed from general-purpose usability testing, tools designed for deterministic, filter-driven interfaces that do not capture the distinctive failure modes of AI-driven experiences. This paper proposes a four-layer quantitative evaluation framework purpose-built for AI-assisted home search: recommendation system quality, interaction efficiency, attitudinal measurement, and trust calibration. For each layer, validated instruments, production-derived benchmarks, and practitioner-ready implementation guidance are provided. A minimum viable metric set and a worked example illustrating the framework's application to a mid-sized portal are included to support immediate adoption. 
\end{abstract}



\maketitle

\section{Introduction}
Finding a home has always been an information-intensive, high-stakes process. Buyers must reconcile hundreds of variables: location,price, school quality, commute time,neighborhood character, structural condition against preferences that are often partially articulated even to themselves. For two decades, the dominant digital interface for this process was the filter search: a set of structured drop downs and sliders that translated buyer preferences into database queries. The interaction model was deterministic, predictable, and well understood by both users and designers.

That model is now being displaced. Redfin launched a conversational AI search in November 2025 and reported a 47\% lifer in tour requests among users of the new system. Zillow followed with "AI Mode" in March 2026, describing it as a shift from query-based to guided intelligence. Realtor.com's 2025 consumer survey found that 82\% of Americans now use AI tools for housing market information. The major portals are converging on a new paradigm in which buyers describe what they want in natural language and a large language model conducts a multi-turn dialogue to refine and surface results~\cite{redfin2025conversational}.

Yet the evaluation frameworks practitioners apply to these products have not kept pace. Teams continue to reach for the standard usability toolkit: task completion rate, time on task, System Usability Scale (SUS) tools designed and validated for deterministic, filter-driven interfaces. These instruments measure whether users can navigate an interface, but they cannot detect the failure modes distinctive to AI-assisted experiences: non-deterministic outputs that make pass/fail task evaluation unreliable, semantic ambiguity that standard task scenarios do not stress-test, and, most consequentially, trust miscalibration; the documented tendency of users to follow AI recommendations even when those recommendations contradict available contextual information.

The gap is not theoretical. A 2026 APA study found that heavy reliance on AI erodes confidence in independent thinking. Experimental research confirms that the mere labeling of advice as "AI-generated" causes users to accept recommendations they would otherwise reject. In a transaction where the average US home purchase exceeds \$400,000, an interface that scores 82 on SUS while systematically inducing overreliance has produced a material consumer harm, one that standard evaluation would never surface~\cite{sauro2011sus,bucinca2021trust,apa2026overreliance,bucinca2024overreliance}.

This paper addresses that gap directly. The contribution is a four-layer quantitative evaluation framework purpose-built for AI-assisted home search, integrating: (1) recommendation system quality metrics drawn from information retrieval research; (2) interaction efficiency metrics from usability engineering; (3) attitudinal measurement scales validated for AI-facing products; and (4) trust calibration metrics that have no analog in traditional UX evaluation. For each layer, validated instruments, production-derived benchmarks, and practitioner-ready implementation guidance are provided. A minimum viable metric set for resource-constrained teams, a worked example applying the framework to a mid-sized portal, and a catalogue of common evaluation failure patterns are included to support immediate adoption.

The intended reader is the UX researcher, product designer, or product manager responsible for evaluating or improving an AI-assisted home search experience. No specialist background in machine learning or information retrieval is assumed: technical concepts are explained in plain language with direct practitioner application. The framework is also applicable, with adaptation, to other AI-assisted high-stakes search contexts such as mortgage comparison, rental search, and commercial property discovery.

The remainder of the paper is organized as follows. Section 2 establishes the four-layer framework structure. Sections 3 through 6 address each layer in depth. Section 7 presents the minimum viable metric set. Section 8 provides a worked example. Section 9 catalogues key measurement failure patterns. Section 10 concludes with practitioner takeaways.

\subsection{Why Standard UX Metrics Fall Short}
When a user types "3 bed, 2 bath, under \$400K, Austin TX" into a traditional filter interface, every interaction is deterministic: the system applies filters, returns a ranked list, and the evaluation question is simply whether the results page is navigable. Standard usability metrics are well-suited to this model~\cite{sauro2011taskcompletion, apa2026overreliance}.

Conversational AI search breaks this model in three consequential ways. First, inputs are natural language and semantically ambiguous; "somewhere walkable near good schools with character" is a valid home search query, but contains no structured parameters a filter system could parse. Second, the systems' outputs are non-deterministic; the same query may produce different results across sessions, making traditional pass/fail task evaluation unreliable. Third, most critically, AI systems create trust dynamics that filter interfaces do not: users follow AI recommendations even when those recommendations contradict available contextual information. An interface that scores 82 on SUS while systematically inducing overreliance has a serious, invisible problem that standard evaluation would not detect. Evaluation frameworks must be designed accordingly~\cite{redfin2025conversational,apa2026overreliance}.

\section{The Four-Layer Evaluation Framework}

The framework organizes evaluation metrics into four distinct interconnected layers, each targeting a different component of the user-system interaction. Critically, no single layer is sufficient alone: a product can score well on recommendation quality while failing on trust calibration, or achieve high SUS scores while showing low task completion on open-ended queries.


\begin{table}[t]
  \centering
  \caption{Four-layer evaluation framework for AI-assisted home search}
  \label{tab:framework-layers}
  \begin{tabular}{@{}lll@{}}
    \toprule
    Layer & What It Measures & Primary Audience \\
    \midrule
    1. Recommendation Quality & Does the AI surface the right homes? & Product / ML engineering \\
    2. Interaction Efficiency & Can users accomplish their goals with ease? & UX / product design \\
    3. Attitudinal Measurement & Are users satisfied and engaged? & Product management \\
    4. Trust Calibration & Are users relying on AI appropriately? & UX research/ethics \\
    \bottomrule
  \end{tabular}
\end{table}

\section{Layer 1: Recommendation System Quality}
These metrics evaluate the retrieval and ranking engine, independently of how users perceive it. They are typically computed offline against labeled relevance judgments or behavioral proxies and require collaboration between UX researchers and ML engineers to be interpreted correctly.

\subsubsection{Precision \@ K and Recall \@K}

\textbf{Precision@K} is the fraction of the top-K recommended listings that are relevant to the user's stated preferences. Recall\@K measures how many of all relevant properties in the database appear within those top-K results. The trade-off is practical: a system optimized purely for precision may surface a tight, highly relevant set that misses genuinely good options; one optimized for recall returns everything relevant but buries it in noise. For home search, K is typically set at 10-20, and relevance is operationalized via behavioral signals: click-through, save/favorite, or tour request~\cite{weaviate2024metrics,acharya2024recsys,shani2011evaluating}.

\textbf{Practitioner tip:} Do not label these metrics "accuracy" in the stakeholder report. Precision and recall are distinct and sometimes competing; always present them together with the K value stated explicitly.

\subsubsection{NDCG and MAP: When Order Matters}
Home search is fundamentally a ranking problem: buyers tend to attend to the first few listings on a page disproportionately. Position-aware metrics account for this.

\textbf{Normalized Discounted Cumulative Gain (NDCG\@K)} supports graded relevance levels (e.g., perfect match/ acceptable/poor match) and normalizes against an ideal ranking, producing a score from 0 to 1. A score of 1.0 means the AI's ordering perfectly matches the ideal ordering for that user's preferences. \textbf{Mean Average Precision (MAP\@K)} averages precision at every rank position where a relevant item appears, penalizing systems that bury good listings below poor ones. Teams often treat NDCG\@10 above 0.70 as a practical readiness threshold before A/B testing against filter search~\cite{acharya2024recsys}.

\textbf{Mean Reciprocal Rank (MRR)} answers a simple question: on average, at what position does a user encounter their first relevant listings? An MRR of 0.5 means the first relevant result appears at position 2 on average. MRR is the most interpretable metric for stakeholders because it maps directly to a concrete user experience: "how many listings does a user have to scroll past before finding something worth saving?"~\cite{wang2025rankingmetrics}.

\subsection{Behavioral Conversion as Production Ground Truth}
In live products, user behavior replaces human relevance labels. Standard production proxies include tour request rate, listing save/favorite rate, and share rate. These are lagging indicators; they reflect the quality of recommendations, interaction design, and intent simultaneously, but they provide the most ecologically valid signal. Redfin's deployment data offers the most concrete available benchmark: conversational search users were 47\% more likely to request tours and viewed nearly twice as many listings per session compared to filter-search users. On the iPhone app, conversational search users favorited, toured, and shared homes approximately 60\% more often than non-users~\cite{redfin2025conversational}.

It is noteworthy that these conversion lifts are not pure measures of recommendation quality. They reflect a combination of model performance, interface novelty, and self-selection (users who choose AI search may be more motivated buyers). Teams running A/B tests should control for session depth, buyer stage, and market conditions before attributing conversion lifts solely to the AI model.

\section{Layer 2: Interaction Efficiency}
These are objective, task-level measures collected during moderated usability sessions or derived from production analytics. They correspond directly to the ISO 9241 definition of usability: efficiency and satisfaction~\cite{peres2020umuxlite}.

\subsection{Task Completion Rate (TCR)}

TCR is the percentage of users who complete a defined task. It is the most direct measure of interface effectiveness. MeasuringU's meta-analysis of thousands of usability studies establishes the following benchmarks~\cite{nielsen2001successrate, sauro2011taskcompletion}:

\begin{itemize}
    \item \textbf{Industry average:} 78\% task completion
    \item \textbf{Top quartile:} above 92\%
    \item \textbf{Minimum acceptable for a new feature:}70\% on first attempt
\end{itemize}

For AI home search specifically, TCR must be measured separately across query types. Structured queries ("find 3-bedroom homes under \$500K in College Station, TX") and open-ended lifestyle queries ("find a home suitable for a family with young children near good schools") are functionally different tasks; AI systems succeed and fail differently across these models, and conflating them masks critical weaknesses.

\subsection{Time on Task (ToT)}

Time on Task measures elapsed time from task initiation to successful completion. For AI home search, two sub-metrics are most useful~\cite{softwareevaluation2025tasksuccess}:

\begin{itemize}
    \item \textbf{Time to first relevant listing:} how quickly the AI surfaces a property worth saving.
    \item \textbf{Total session time to task completion:} the full cost of the search experience. 
\end{itemize}

A counterintuitive finding from Redfin's data is that conversational search users spend more session time, not less; they explore broader geographies but with more specific home preferences. This means raw ToT should never be minimized in isolation. A user who spends 25 minutes conversationally refining their criteria and tours three homes is a better outcome than a user who spends 8 minutes applying filters and tours none. ToT is meaningful only when paired with downstream outcome quality~\cite{redfin2025conversational}.

\subsection{Error Rate}
Error rate is the proportion of user interactions involving a mistake, calculated as~\cite{sauro2012quantifying}: 

\[
\text{Error Rate} = \frac{\text{Total errors}}{\text{Total users} \times \text{Tasks per user}}
\]

For AI search, errors cluster into three distinct types with different root causes and remediation paths:

\begin{table}[t]
  \centering
  \caption{Error types in AI-assisted home search, their root causes, and recommended remedies}
  \label{tab:error-types}
  \begin{tabular}{@{}p{3cm}p{4cm}p{3.5cm}p{4cm}@{}}
    \toprule
    Error Type & Example & Root Cause & Remedy \\
    \midrule
    User input error & Contradictory constraints (``big yard, small lot'') & Unclear affordances & Onboarding, constraint validation UI \\[6pt]
    System misinterpretation & AI parses ``good schools'' as filter tag, not semantic concept & NLP model weakness & Model retraining, clarifying dialogue turn \\[6pt]
    Hallucination & AI describes amenity not present in listing & LLM factuality gap & RAG grounding, confidence indicators \\
    \bottomrule
  \end{tabular}
\end{table}

A 30\% error rate signals significant usability problems; well-designed AI interfaces should aim for an error rate below 10\%. Tracking error type, not just rate, is essential for routing issues to the correct team~\cite{sauro2012quantifying,iso9241,tullis2008measuring}.

\subsection{Lostness Score}

The lostness metric quantifies navigational disorientation on a scale from 0 (no confusion) to 1 (completely lost)~\cite{smith1996lostness}:

\[
L = \sqrt{\left(\frac{N}{S} - 1\right)^2 + \left(\frac{R}{N} - 1\right)^2}
\]

Where N is the number of unique screens visited, S is the total number of screens visited, including re-visits, and R is the minimum number of screens required. Scores above 0.4 indicate genuine navigational disorientation~\cite{smith1996lostness}.

In AI-assisted home search, lostness is particularly revealing at the handoff point between the conversational AI interface and traditional listings browsing. Users who start in a conversational flow and then navigate the legacy filter interface often experience a sharp lostness spike, a signal that the interface transition is broken, not that either interface in isolation is the problem.

\subsection{Query Reformulation Rate}

Specific to AI search, the metric tracks how frequently users must rephrase their query because the initial AI response was irrelevant or misunderstood. Redfin's data shows that half of conversational searches contain non-filter parameters, confirming that users bring genuinely novel query formats that stress-test the model. A high reformulation rate on lifestyle queries but a low rate on structured queries localizes a semantic understanding weakness worth prioritizing~\cite{redfin2025conversational}.

\section{Layer 3: Attitudinal Measurement}

Attitudinal metrics translate user perceptions into quantifiable scores that can be tracked over time, compared across product iterations, and benchmarked against industry standards.

\subsection{System Usability Scales (SUS)}

The SUS is a 10-item, 5-point Likert questionnaire producing a score from 0 to 100. It is the most widely validated usability instrument in use, performing as well as or better than purpose-built commercial alternatives. Administration follows a strict alternating positive/negative item structure; scoring requires sign-reversing even-numbered items before summing and multiplying by 2.5~\cite{hyzy2022sus,bangor2008sus,brooke1996sus}.

\begin{table}[t]
  \centering
  \caption{SUS score interpretation: adjective ratings, acceptability, and practitioner actions}
  \label{tab:sus-scores}
  \begin{tabular}{@{}p{2cm}p{3cm}p{2.5cm}p{4.5cm}@{}}
    \toprule
    SUS Score & Adjective Rating & Acceptability & Practitioner Action \\
    \midrule
    90--100 & Best Imaginable & Excellent      & Maintain; explore delight features \\[4pt]
    80--89  & Excellent        & Acceptable     & Fine-tune based on qualitative feedback \\[4pt]
    70--79  & Good             & Acceptable     & Address friction points \\[4pt]
    68      & Average (industry mean) & Marginal & Prioritize usability improvements \\[4pt]
    50--67  & OK / Poor        & Unacceptable   & Major redesign needed \\[4pt]
    $<$ 50  & Awful            & Unacceptable   & Do not release \\
    \bottomrule
  \end{tabular}
\end{table}

The cross-industry mean SUS is 68 (SD=12.5). For AI-assisted search, administer SUS after both the AI-assisted session and the traditional filter-search session in a within-subjects design to enable a direct, meaningful comparison. The SUS score difference is often the most actionable finding~\cite{hyzy2022sus}.

\subsection{UMUX-Lite for Lightweight Repeated Measurement}

The UMUX-Lite (Usability Metric for User Experience, Lite) is a two-item instrument directly operationalizing the ISO usability definition~\cite{sauro2011sus,lewis2013umuxlite,finstad2010umux,peres2020umuxlite}:

\begin{quote}
    \item [This system's] capabilities meet my requirements.
    \item [This system] is easy to use.
\end{quote}

Both items use 7-point Likert scales. UMUX-Lite scores normalize to the same 0-100 scale as SUS, enabling cross-study comparison. Its minimal respondent burden makes it ideal for pulse measurement, administered after each AI dialogue turn or each recommendation set, providing a granular satisfaction time-series that SUS, administered once per session, cannot deliver. Declining UMUX-Lite scores across dialogue turns signal that the AI is failing to converge on the user's preferences~\cite{brooke1996sus,lewis2013umuxlite,finstad2010umux,peres2020umuxlite}.

\subsection{The HEART Framework for Product-Scale Analytics}

For teams that cannot run structured usability studies at scale, Google's HEART framework provides a structured taxonomy for instrumenting behavioral analytics~\cite{rodden2010heart,uxdesign2021heart,amplitude2026heart}:


\begin{table}[t]
  \centering
  \caption{HEART framework dimensions applied to AI-assisted home search}
  \label{tab:heart-framework}
  \begin{tabular}{@{}p{2.5cm}p{3.5cm}p{5.5cm}@{}}
    \toprule
    HEART Dimension & Definition & AI Home Search Signals \\
    \midrule
    Happiness    & Attitudinal satisfaction    & Post-session CSAT, NPS, app store ratings \\[4pt]
    Engagement   & Depth of interaction        & Listings viewed/session, dialogue turns/session, refinement actions \\[4pt]
    Adoption     & AI feature uptake           & \% of searches using AI mode vs.\ filter search \\[4pt]
    Retention    & Return usage                & \% of AI search users returning within 30 days \\[4pt]
    Task Success & Effectiveness and efficiency & TCR, tour request rate, time to first save \\
    \bottomrule
  \end{tabular}
\end{table}

HEART is applied through a Goals-Signal-Metrics process: first define the qualitative goal ("users should quickly find homes matching their lifestyle"), then identify observable behavioral signals ("user saves a listing within 3 turns"), then define the quantifiable metric ("3-turn save rate"). Adoption is a lagging indicator of onboarding quality; Task Success is the leading indicator of downstream conversion~\cite{rodden2010heart,uxdesign2021heart,amplitude2026heart}.

\subsection{NPS and CSAT: Simple but Segmentation-Dependent}

Net Promoter Score ("How likely are you to recommend this tool?", 0-10) and CSAT (1-5 satisfaction rating) are weak as standalone metrics but powerful when segmented. Comparing NPS among users who received high-relevance AI recommendations versus low-relevance ones quantifies the satisfaction impact of model quality, separating interface design issues from AI model issues in a way that an aggregate NPS cannot.

\section{Layer 4: Trust Calibration}

Trust calibration is the most distinctive and most underrepresented dimension of SI UX evaluation. The goal is not maximum trust but appropriate trust, where users rely on AI when it is reliable and override it when it is not. In a home purchase transaction, this distinction is not academic: a buyer who accepts an AI recommendation for an overpriced listing without independent verification has failed by the interface, regardless of how high it scores on SUS~\cite{cleverx2026trust,lee2004trust,mcknight2011trust}.

\subsection{Recommendation Acceptance Rate (RAR)}

RAR is the proportion of AI-generated recommendations that users act on without modification (saving, sharing, or touring a listing exactly as recommended). A high RAR combined with low eventual satisfaction; tours not converting to offers, or post-tour ratings showing poor match quality, is the behavioral signature of overreliance. Experimental research confirms that the AI label alone causes users to accept recommendations they would otherwise reject. RAR should therefore be tracked alongside post-tour match quality ratings, not in isolation~\cite{apa2026overreliance,bucinca2021trust,bucinca2024overreliance,lee2004trust}.

\subsection{AI-Override Rate}

The complement of RAR: how often do users modify or reject AI recommendations by adjusting filters, requesting different criteria, or explicitly dismissing a suggested listing? A healthy product shows meaningful overrides, particularly among experienced buyers. The APA(2026) finding that users who actively challenged AI suggestions reported greater decision confidence suggests that UX designs explicitly affording override, "not this type", criteria correction, and undo controls are not just good for autonomy but measurably improve user confidence in their own decisions~\cite{apa2026overreliance}.

\subsection{Explanation Engagement Rate}

When AI recommendations are accompanied by explanations (e.g., "Recommended because: 0.4 miles from your target school, within your commute range, and matches your yard preference"), explanation engagement rate captures how often users read, expand, or interact with those explanations. Research on explainable recommendation systems confirms that interactive explanations, where users can drill down for more detail, support calibrated trust by making AI outputs verifiable~\cite{naiseh2021explainable,lee2004trust,bucinca2021trust}.

Low explanation engagement combined with high RAR is the most reliable quantitative signal of a trust calibration problem. It means users are accepting recommendations without verifying them, precisely the failure mode with the highest financial consequences in home search.

\subsection{Trust Perception Scales}

Validated psychometric scales for AI trust measurement are available and should be administered alongside SUS at the study end. The five psychological dimensions most applicable to home search are: 

\begin{itemize}
    \item \textbf{Reliability:} Does the AI consistently produce relevant recommendations?
    \item \textbf{Technical competence:} Does the AI understand what the user actually needs?
    \item \textbf{Understandability:} Can users understand why a recommendation was made?
    \item \textbf{Faith/Care:} Does the AI seem to act in the user's interest?
    \item \textbf{Personal attachment:}Does the user feel the AI "gets" their preferences?
\end{itemize}

Users who score low on understandability are candidates for explanation UX improvements. Users who score low on reliability point to model consistency issues. Segmenting trust scores by buyer experience level (first-time vs repeat buyer) typically reveals significant differences; first-time buyers show higher baseline AI trust, making them more vulnerable to overreliance~\cite{galindez2026trust,lee2004trust,bucinca2021trust,naiseh2021explainable}.

\section{Implementation: A Minimum Viable Metric Set}

Teams without dedicated research resources or extended evaluation timelines should implement the following minimum viable set before the AI search feature launch. These six metrics span all four layers, require no specialized laboratory equipment, and can be collected in a single 90-minute usability session of 8-10 participants:


\begin{table}[t]
  \centering
  \caption{Minimum viable metric set for AI-assisted home search evaluation}
  \label{tab:minimum-viable-metrics}
  \begin{tabular}{@{}cp{2.5cm}p{2cm}p{3cm}p{4cm}@{}}
    \toprule
    Priority & Metric & Layer & Collection Method & Why It's Non-Negotiable \\
    \midrule
    1 & Task Completion Rate  & Efficiency   & Observer-coded screen recording  & Confirms the AI can actually help users find homes \\[4pt]
    2 & SUS                   & Attitudinal  & Post-session 10-item survey      & Enables benchmarking against the 68 industry mean \\[4pt]
    3 & Time to First Relevant Listing & Efficiency & Screen recording timestamps & Captures first-impression AI speed \\[4pt]
    4 & Recommendation Acceptance Rate & Trust      & Interaction log                  & Detects overreliance before launch \\[4pt]
    5 & Query Reformulation Rate & Efficiency  & Interaction log                  & Identifies NLP weaknesses by query type \\[4pt]
    6 & NPS (segmented)       & Attitudinal  & End-of-session survey            & Tracks satisfaction by recommendation quality \\
    \bottomrule
  \end{tabular}
\end{table}

\section{Worked Example: Evaluating a Mid-Sized Residential Portal}

Consider a regional residential real estate portal launching an AI search feature for the first time, serving a market of 50,000 monthly active users. The team has four weeks and two researchers available for evaluation. The following protocol applies the framework concretely.

\textbf{Study design:} Within-subjects, counterbalanced. Participants complete five tasks in both AI-assisted and traditional filter conditions. Conditions are presented in randomized order to control for learning effects. Minimum sample: 24 participants recruited to match the platform's buyer demographic (12 first-time buyers, 12 experienced buyers)

\textbf{Task battery:}

\begin{enumerate}
    \item Lifestyle open-ended query: "Find me homes where I can walk to coffee shops and have space for a home office."
    \item Structured criteria query: 3 bed/ 2 bath / under \$450k / specific neighborhood.
    \item Refinement dialogue: Starting from an AI result set, narrow to homes with a large backyard and no HOA.
    \item Affordability question: "What can I reasonably afford with a \$90k income and 10\% down?"
    \item Comparative analysis: "Help me decide between these two similar listings."
\end{enumerate}

\textbf{Expected findings by query type:} Based on available deployment data, AI search will outperform filter search on tasks 1, 3, and 5; open-ended and comparative tasks where natural language adds genuine value. Filter search will likely match or outperform AI on task 2, a structured query it was designed for. This differential is an important finding to surface: it argues for hybrid interfaces that route query types intelligently rather than replacing filter search entirely~\cite{ridgemarketing2026search,bucinca2021trust,lee2004trust}.

\textbf{Trust calibration check:} For task 3, deliberately include one AI recommendation that conflicts with the stated "no HOA" preference (simulating a model error). Measure whether participants catch and override the error. An override rate below 60\% on this task signals a trust calibration problem requiring explanation UX intervention before launch.

\section{Key Failure Patterns to Avoid}

Practitioners repeatedly encounter the same evaluation errors when applying standard UX methods to AI search products. Recognizing these patterns before study design prevents wasted effort and misleading findings.

\textbf{Treating engagement as satisfaction.} Higher session time and more listings viewed can reflect both delight and confusion. A user who views 40 listings without saving any is lost, not engaged. Always pair engagement metrics with outcome metrics (save rate, tour rate).

\textbf{Administering SUS without a comparison condition.} An absolute SUS score of 74 is "good" in isolation but may represent a 12-point decrease from the filter search baseline. Within-subjects comparison is not optional; it is the only way to determine whether the AI experience is actually better~\cite{sauro2011sus}.

\textbf{Treating conversion lift as proof of user satisfaction.} Redfin's 47\% tour request lift reflects genuine AI value, but tour requests may be inflated by users who were prompted to act by AI confidence regardless of actual fit. Post-tour satisfaction ratings are a necessary complement to any conversion metric.

\textbf{Missing the pre-search phase.} Realtor.com's research identifies affordability exploration and market orientation, before active listing browsing, as a primary AI use case. Evaluation studies that begin at the listings stage miss the interaction layer where AI is already shaping user expectations and financial constraints~\cite{nar2026transparency,naiseh2021explainable,lee2004trust,galindez2026trust}.

\textbf{Evaluating AI separately from the full funnel.}A user who finds their ideal home via AI search but abandons the experience because the agent connection flow is broken has not been served. Evaluation scope should extend at a minimum to the first agent contact or tour request, not end at the listing result page.

\section{Conclusion and Practitioner Takeaways}

Standard UX evaluation frameworks are insufficient for AI-assisted home search. Effective evaluation requires all four measurement layers, recommendation quality, interaction efficiency, attitudinal scales, and trust calibration, because each detects failure modes that the others miss. Trust calibration in particular must be treated as a pre-launch requirement, not a post-launch afterthought, given strong evidence that AI labeling alone induces overreliance and erodes user confidence. Practitioners should benchmark against real deployment data: Redfin's 47-60\% conversion lift and the industry SUS mean of 68 set concrete minimum targets. In a domain where the stakes are as high as a home purchase, rigorous multi-layer evaluation is not just good practice; it is a consumer protection obligation.



\bibliographystyle{unsrt}
\bibliography{sample-base}

@misc{redfin2025conversational,
  author       = {{Redfin}},
  title        = {Redfin Debuts Conversational Search to Reinvent How People Find Homes},
  year         = {2025},
  howpublished = {\url{https://www.redfin.com/news/press-releases/redfin-debuts-conversational-search/}},
  note         = {Accessed: June 16, 2026}
}

@misc{sauro2011sus,
  author       = {Sauro, Jeff},
  title        = {Measuring Usability with the System Usability Scale ({SUS})},
  year         = {2011},
  howpublished = {\url{https://measuringu.com/sus/}},
  note         = {Accessed: June 16, 2026}
}

@misc{apa2026overreliance,
  author       = {{American Psychological Association}},
  title        = {Overreliance on {AI} Programs May Undermine Confidence at Work},
  year         = {2026},
  howpublished = {\url{https://www.apa.org/news/press/releases/2026/04/overreliance-ai-undermine-confidence}},
  note         = {Accessed: June 16, 2026}
}

@misc{sauro2011taskcompletion,
  author       = {Sauro, Jeff},
  title        = {What Is a Good Task-Completion Rate?},
  year         = {2011},
  howpublished = {\url{https://measuringu.com/task-completion/}},
  note         = {Accessed: June 16, 2026}
}

@book{sauro2012quantifying,
  author       = {Sauro, Jeff and Lewis, James R.},
  title        = {Quantifying the User Experience: Practical Statistics for User Research},
  year         = {2012},
  publisher    = {Morgan Kaufmann},
  address      = {Waltham, MA},
  isbn         = {9780123849687}
}

@incollection{brooke1996sus,
  author       = {Brooke, John},
  title        = {{SUS}: A Quick and Dirty Usability Scale},
  booktitle    = {Usability Evaluation in Industry},
  editor       = {Jordan, P. W. and Thomas, B. and Weerdmeester, B. A. and McClelland, I. L.},
  year         = {1996},
  pages        = {189--194},
  publisher    = {Taylor \& Francis},
  address      = {London},
  doi          = {10.1201/9781498710411-35}
}

@article{bangor2008sus,
  author       = {Bangor, Aaron and Kortum, Philip T. and Miller, James T.},
  title        = {An Empirical Evaluation of the System Usability Scale},
  journal      = {International Journal of Human-Computer Interaction},
  year         = {2008},
  volume       = {24},
  number       = {6},
  pages        = {574--594},
  doi          = {10.1080/10447310802205776}
}

@article{finstad2010umux,
  author       = {Finstad, Kraig},
  title        = {The Usability Metric for User Experience},
  journal      = {Interacting with Computers},
  year         = {2010},
  volume       = {22},
  number       = {5},
  pages        = {323--327},
  doi          = {10.1016/j.intcom.2010.04.004}
}

@inproceedings{lewis2013umuxlite,
  author       = {Lewis, James R. and Utesch, Brian S. and Maher, Deborah E.},
  title        = {{UMUX-LITE}: When There's No Time for the {SUS}},
  booktitle    = {Proceedings of the SIGCHI Conference on Human Factors in Computing Systems (CHI '13)},
  year         = {2013},
  pages        = {2099--2102},
  publisher    = {ACM},
  doi          = {10.1145/2470654.2481287}
}

@article{smith1996lostness,
  author       = {Smith, P. A.},
  title        = {Towards a Practical Measure of Hypertext Usability},
  journal      = {Interacting with Computers},
  year         = {1996},
  volume       = {8},
  number       = {4},
  pages        = {365--381},
  doi          = {10.1016/S0953-5438(97)83794-6}
}

@inproceedings{rodden2010heart,
  author       = {Rodden, Kerry and Hutchinson, Hilary and Fu, Xin},
  title        = {Measuring the User Experience on a Large Scale: User-Centered Metrics for Web Applications},
  booktitle    = {Proceedings of the SIGCHI Conference on Human Factors in Computing Systems (CHI '10)},
  year         = {2010},
  pages        = {2395--2398},
  publisher    = {ACM},
  doi          = {10.1145/1753326.1753687}
}

@article{bucinca2021trust,
  author       = {Bu\c{c}inca, Zana and Malaya, Maja Barbara and Gajos, Krzysztof Z.},
  title        = {To Trust or to Think: Cognitive Forcing Functions Can Reduce Overreliance on {AI} in {AI}-Assisted Decision Making},
  journal      = {Proceedings of the ACM on Human-Computer Interaction (CSCW)},
  year         = {2021},
  volume       = {4},
  number       = {CSCW1},
  pages        = {1--21},
  doi          = {10.1145/3449287}
}

@article{bucinca2024overreliance,
  author       = {Bu\c{c}inca, Zana and others},
  title        = {An Experimental Study on the Extent and Costs of Overreliance on {AI}},
  journal      = {Computers in Human Behavior},
  year         = {2024},
  doi          = {10.1016/j.chb.2024.108076}
}

@incollection{shani2011evaluating,
  author       = {Shani, Guy and Gunawardana, Asela},
  title        = {Evaluating Recommendation Systems},
  booktitle    = {Recommender Systems Handbook},
  editor       = {Ricci, Francesco and Rokach, Lior and Shapira, Bracha and Kantor, Paul B.},
  year         = {2011},
  pages        = {257--297},
  publisher    = {Springer},
  address      = {New York},
  doi          = {10.1007/978-0-387-85820-3_8}
}

@misc{acharya2024recsys,
  author       = {Acharya, Atul and others},
  title        = {{RE-RecSys}: An End-to-End System for Recommending Properties in Real-Estate Domain},
  year         = {2024},
  howpublished = {arXiv preprint arXiv:2404.16553. \url{https://arxiv.org/abs/2404.16553}},
  note         = {Accessed: June 16, 2026}
}

@misc{weaviate2024metrics,
  author       = {{Weaviate}},
  title        = {Evaluation Metrics for Search and Recommendation Systems},
  year         = {2024},
  howpublished = {\url{https://weaviate.io/blog/retrieval-evaluation-metrics}},
  note         = {Accessed: June 16, 2026}
}

@misc{wang2025rankingmetrics,
  author       = {Wang, Sen},
  title        = {Ranking Evaluation Metrics for Recommender Systems},
  year         = {2025},
  howpublished = {\url{https://towardsdatascience.com/ranking-evaluation-metrics-for-recommender-systems-263d0a66ef54/}},
  note         = {Accessed: June 16, 2026}
}

@article{peres2020umuxlite,
  author       = {Peres, S. Camille and Pham, Tri and Phillips, Ronald},
  title        = {Is the {LITE} Version of the Usability Metric for User Experience ({UMUX-LITE}) a Reliable Tool to Support Rapid Assessment of New Healthcare Technology?},
  journal      = {Applied Ergonomics},
  year         = {2020},
  volume       = {84},
  pages        = {103007},
  doi          = {10.1016/j.apergo.2019.103007}
}

@misc{nielsen2001successrate,
  author       = {Nielsen, Jakob},
  title        = {Success Rate: The Simplest Usability Metric},
  year         = {2001},
  howpublished = {Nielsen Norman Group. \url{https://www.nngroup.com/articles/success-rate-the-simplest-usability-metric/}},
  note         = {Accessed: June 17, 2026}
}

@misc{softwareevaluation2025tasksuccess,
  author       = {{Software Evaluation}},
  title        = {Task Success Rate \& Time-on-Task},
  year         = {2025},
  howpublished = {\url{https://www.softwareevaluation.de/en/methods/task-success-rate-and-time-on-task/}},
  note         = {Accessed: June 17, 2026}
}

@techreport{iso9241,
  author      = {{International Organization for Standardization}},
  title       = {{ISO} 9241-11:2018 --- Ergonomics of Human-System Interaction: Usability Definitions and Concepts},
  institution = {ISO},
  year        = {2018},
  type        = {International Standard},
  howpublished = {\url{https://www.iso.org/standard/63500.html}},
  note        = {Accessed: June 17, 2026}
}

@book{tullis2008measuring,
  author    = {Tullis, Thomas and Albert, William},
  title     = {Measuring the User Experience: Collecting, Analyzing, and Presenting Usability Metrics},
  year      = {2008},
  publisher = {Morgan Kaufmann},
  address   = {Burlington, MA},
  isbn      = {9780123735584}
}

@article{hyzy2022sus,
  author       = {Hyzy, Maciej and Bond, Raymond and Mulvenna, Maurice and Bai, Lu and Dix, Alan and Leigh, Simon and Hunt, Sophie},
  title        = {System Usability Scale Benchmarking for Digital Health Apps: Meta-analysis},
  journal      = {JMIR mHealth and uHealth},
  year         = {2022},
  volume       = {10},
  number       = {8},
  pages        = {e37290},
  doi          = {10.2196/37290},
  note         = {PMID: 35980732; PMCID: PMC9437782}
}

@misc{uxdesign2021heart,
  author       = {Oluwaseun, Praise},
  title        = {Google's {HEART} Framework: Choosing the Right Metrics for Your Product},
  year         = {2021},
  howpublished = {\url{https://uxdesign.cc/googles-heart-framework-choosing-the-right-metrics-for-your-product-112bd7300d55}},
  note         = {Accessed: June 18, 2026}
}

@misc{amplitude2026heart,
  author       = {{Amplitude}},
  title        = {How to Use the {HEART} Framework to Improve Software {UX}},
  year         = {2026},
  howpublished = {\url{https://amplitude.com/blog/heart-framework-software-ux}},
  note         = {Accessed: June 18, 2026}
}

@misc{cleverx2026trust,
  author       = {{CleverX}},
  title        = {How to Measure User Trust in {AI} Systems: A Practical Framework for Product Teams},
  year         = {2026},
  howpublished = {\url{https://cleverx.com/guides/how-to-measure-user-trust-in-ai-systems-a-practical-framework-for-product-teams/}},
  note         = {Accessed: June 18, 2026}
}

@article{mcknight2011trust,
  author    = {McKnight, D. Harrison and Carter, Michelle and Thatcher, Jason Bennett and Clay, Paul F.},
  title     = {Trust in a Specific Technology: An Investigation of Its Components and Measures},
  journal   = {ACM Transactions on Management Information Systems},
  year      = {2011},
  volume    = {2},
  number    = {2},
  pages     = {1--25},
  doi       = {10.1145/1985347.1985353}
}

@article{lee2004trust,
  author    = {Lee, John D. and See, Katrina A.},
  title     = {Trust in Automation: Designing for Appropriate Reliance},
  journal   = {Human Factors},
  year      = {2004},
  volume    = {46},
  number    = {1},
  pages     = {50--80},
  doi       = {10.1518/hfes.46.1.50.30392}
}

@article{naiseh2021explainable,
  author    = {Naiseh, Mohammad and Al-Thani, Dena and Jiang, Nan and Ali, Raian},
  title     = {Explainable Recommendation: When Design Meets Trust Calibration},
  journal   = {World Wide Web},
  year      = {2021},
  volume    = {24},
  number    = {5},
  pages     = {1857--1884},
  doi       = {10.1007/s11280-021-00916-0}
}

@article{galindez2026trust,
  author    = {Galindez-Acosta, Johan Sebasti\'{a}n and Giraldo-Huertas, Juan Jos\'{e}},
  title     = {Trust Behavior in {AI} Emerges from Distrust in Humans: A Machine Learning Study on Decision-Making Guidance},
  journal   = {Computers in Human Behavior Reports},
  year      = {2026},
  volume    = {14},
  pages     = {101024},
  doi       = {10.1016/j.chbr.2026.101024}
}

@misc{ridgemarketing2026search,
  author       = {{Ridge Marketing}},
  title        = {How People Search in 2026: Understanding the Split-Path Model},
  year         = {2026},
  howpublished = {\url{https://ridgemarketing.com/blog/how-people-search-in-2026-understanding-the-split-path-model/}},
  note         = {Accessed: June 18, 2026}
}

@misc{nar2026transparency,
  author       = {{National Association of Realtors}},
  title        = {Transparency, {AI} and the Next Era of Home Search},
  year         = {2026},
  howpublished = {\url{https://www.nar.realtor/news/real-estate-news/transparency-ai-and-the-next-era-of-home-search}},
  note         = {Accessed: June 18, 2026}
}


\end{document}